\documentclass[aps,JCAP,onecolumn,superscriptaddress, preprintnumbers]{revtex4}

\usepackage{graphicx}
\usepackage{amsmath}
\usepackage{epsfig, epstopdf}
\usepackage{amsmath,amsfonts,amssymb}
\usepackage{color}
\usepackage{color}

\usepackage{natbib}
\begin{document}
\baselineskip=12pt
\def\black{\textcolor{black}}
\def\red{\textcolor{black}}
\def\blue{\textcolor{blue}}
\def\green{\textcolor{black}}
\def\beq{\begin{equation}}
\def\eeq{\end{equation}}
\def\bea{\begin{eqnarray}}
\def\eea{\end{eqnarray}}
\def\orc{\Omega_{r_c}}
\def\om{\Omega_{\text{m}}}
\def\E{{\rm e}}
\def\bearst{\begin{eqnarray*}}
\def\eearst{\end{eqnarray*}}
\def\peleven{\parbox{11cm}}
\def\peffec{\peight{\bearst\eearst}\hfill\peleven}
\def\pspace{\peight{\bearst\eearst}\hfill}
\def\ptwelve{\parbox{12cm}}
\def\peight{\parbox{8mm}}

\newcommand{\nc}{\newcommand}
\nc{\ba}{\begin{eqnarray}}
\nc{\ea}{\end{eqnarray}}


\title{Large Scale Anisotropic Bias from Primordial non-Gaussianity}

\author{Shant Baghram}
\email{baghram-AT-ipm.ir}
\address{School of Astronomy, Institute for Research in
Fundamental Sciences (IPM),
P.~O.~Box 19395-5531,
Tehran, Iran}
\author{Mohammad Hossein Namjoo}
\email{mh.namjoo-AT-ipm.ir}

\address{School of Physics, Institute for Research in
Fundamental Sciences (IPM),
P.~O.~Box 19395-5531,
Tehran, Iran}

\author{Hassan Firouzjahi}
\email{firouz-AT-ipm.ir}

\address{School of Astronomy, Institute for Research in
Fundamental Sciences (IPM),
P.~O.~Box 19395-5531,
Tehran, Iran}

\vskip 1cm

\begin{abstract}
In this work we study the large scale structure bias in models of anisotropic inflation. We use the Peak Background Splitting method in Excursion Set Theory to find the scale-dependent bias. We show that the amplitude of the bias is modified by a direction-dependent factor. In the specific anisotropic inflation model which we study, the scale-dependent bias vanishes at leading order when the long wavelength mode in squeezed limit is aligned with the anisotropic direction in the sky. We also extend the scale-dependent bias formulation to the  general situations with  primordial anisotropy.  We find some selection rules indicating that some specific parts of a generic anisotropic bispectrum is picked up by the bias parameter. We argue that the anisotropic bias is mainly sourced by the angle between the anisotropic direction and the long wavelength mode in the squeezed limit.

\vskip 0.5cm
\end{abstract}

\preprint{ IPM/A-2013/23}

\maketitle



\section{INTRODUCTION}

Inflation \cite{Guth:1980zm} has emerged as the leading paradigm for the theory of early Universe and structure formation. Basic predictions of inflation indicate that the curvature perturbations are nearly scale-invariant, nearly adiabatic and nearly Gaussian which are in very good agreements with cosmological observations such as WMAP \cite{Hinshaw:2012fq} and
PLANCK \cite{Ade:2013zuv}.
 The simplest models of  inflation are based on a scalar field rolling slowly on a flat potential.
 Any detection of primordial non-Gaussianity (NG) will have significant implications for inflationary model buildings, for a review see \cite{Chen:2010xka, Komatsu:2010hc, Bartolo:2010qu}. For example, many models of single field  inflation predict a very small amount of local non-Gaussianity in the squeezed limit $f_{NL} \sim (1-n_s)$ \cite{Maldacena:2002vr},  in which $n_s$ is the curvature perturbation power spectrum spectral index and $f_{NL}$ parametrizes the amplitude of local NG. With $n_s \simeq 0.96$
from PLANCK \cite{Ade:2013zuv}, one expects $f_{NL} \sim {\cal {O}} (10^{-2})$ for conventional models of
single field inflation. However, this expectation is violated if the system has not reached the attractor regime \cite{Namjoo:2012aa, Chen:2013aj} or if one allows for a non Bunch-Davies
initial condition \cite{Agullo:2010ws, Ashoorioon:2010xg,  Ganc:2011dy, Chialva:2011hc}. Furthermore, inflationary models with large NG predicts different shapes for bispectrum. Therefore, any detection or otherwise of large primordial NG with different shapes will go a long way to rule out many inflationary scenarios or put constraints on model parameters. Having this said, the recent PLANCK collaboration data \cite{Ade:2013ydc} showed no significant deviation from Gaussian initial conditions. PLANCK constrained the amplitude of NG for different known shapes and accordingly the Gaussian initial conditions are consistent with the picture.

The most suitable cosmological observation to constrain the primordial NG is CMB. This is because the perturbations in the last scattering surface are in the linear regime and the fingerprints of non-Gaussianity are mainly
preserved \cite{Komatsu:2001wu,Komatsu:2003iq}. However, recently the interests in Large Scale Structure (LSS) observations and their implications for non-Gaussianity are boosted due to the theoretical findings of  scale-dependent bias \cite{Dalal:2007cu}.  In general,  the distribution of baryonic matter in the Universe, mainly clustered  in galaxies and the clusters of galaxies, is the fundamental observable of LSS \cite{Scoccimarro:2003wn,Desjacques:2010nn}. The distribution of galaxies and clusters of galaxies can be studied by a): the mass function of the structures (i.e. galaxies and cluster of galaxies) and b): the correlation functions, power spectrum and even higher moments of distribution. In the standard theories of structure formation, the  Gaussianity assumption plays a crucial role in finding the  distribution of structures via the primordial density contrast distribution \cite{Press:1973iz}. Accordingly, changing the initial condition from a Gaussian to
non-Gaussian primordial density perturbations will change the  mass function of structures. This change mainly shows itself in the tail of distribution function. Consequently, this effect manifests itself mainly in the statistics of the clusters of galaxies in high mass and high redshifts distributions \cite{Verde:2010wp,Matarrese:2000iz,Kamionkowski:2008sr,LoVerde:2007ri,Grossi:2009an}.
The bispectrum of LSS observations is also affected by the primordial NG and by the secondary NG induced by the non-linear growth of structures
 \cite{Catelan1995}. 
The NG introduced by non-linear growth of structures has its own signatures on the bispectrum of structures, but its shape changes in deep non-linear regime. One can find  the bispectrum of galaxies and clusters of galaxies to detect the primordial non-Gaussianity and distinguish its effect from the effects of gravitational instability. Recent works have shown that the galaxy bispectrum can be a  very promising way to constrain primordial NG \cite{Scoccimarro:2003wn, Jeong:2009vd, Nishimichi:2009fs, Baldauf:2010vn}. On the other hand,  the primordial NG may have an effect on the clustering of halos. As an intuitive example we can consider the local NG where the long wavelength mode  changes the background linear density fluctuations. This change, in the picture of linear bias theory, will have an influence on  the density peaks where structures are formed. In other words, the non-Gaussian long mode  changes the threshold which structure goes from linear to non-linear regime.

The primordial NG has  unique feature in LSS by introducing a scale-dependent bias \cite{Dalal:2007cu, Matarrese:2008nc,Afshordi:2008ru}. This  fingerprint of primordial NG provides the opportunity to constrain primordial NG using the power spectrum of galaxies. Many works are done in numerical simulations to check the scale-dependence of the bias parameter introduced by primordial NG and also to study the change in the statistics of structures in Universe \cite{Grossi:2009an,Reid:2010vc,Sefusatti:2011gt}. From the observational side different groups used the LSS probes to constrain the primordial
NG \cite{Xia:2011hj,Pullen:2012rd,Giannantonio:2013uqa}. Furthermore,  there are other LSS observables such as the Integrated Sachs Wolfe cross correlation with the galaxy power spectrum \cite{Giannantonio:2013uqa}, the 3D bispectrum of Ly-alpha forest,  the redshifted 21-cm signal from the post re-ionization epoch \cite{Sarkar:2012mq,Hazra:2012qz}, the statistics of voids \cite{D'Amico:2010kh}, the cosmological weak lensing \cite{Fedeli:2012dg} etc. which can also be used to study the primordial NG.  In a word, the LSS observations will become a very important tool, complementary to CMB observations, to constrain the properties of primordial NG.

There have been some indications of statistical anisotropies on CMB power spectrum. Although
The statistical significance of the violation of statistical isotropy is not high in WMAP data \cite{Hanson:2009gu, Hanson:2010gu}, but nonetheless the possibility of having statistically anisotropic seed perturbations are intriguing. Recently the data from PLANCK collaboration also confirmed  the anomalies observed by WMAP, including the anisotropy in CMB sky \cite{Ade:2013nlj}. This observation  triggers the interests in anisotropic models both theoretically and observationally.

The statistical anisotropy is usually parameterized via \cite{Ackerman:2007nb} ${\cal P}_\zeta = {\cal P_\zeta}_{0} (1+ g_* (\hat k. \hat n)^2 )$ in which $\hat n$ is the preferred direction in sky, ${\cal P}_\zeta$ is the curvature perturbation power spectrum of the Fourier mode $\vec{k}$ with the direction along the unit vector $\hat k$ and ${\cal P_\zeta}_{0}$ represents the isotropic power spectrum.
Constraints from CMB and large scale structure indicate that $| g_*| \lesssim 0.4$ \cite{Ade:2013ydc, Groeneboom:2009cb, Pullen:2010zy}. Recently the bispectrum and the trispectrum in a model of anisotropic inflation \cite{Watanabe:2009ct} have been calculated in
\cite{Bartolo:2012sd, Shiraishi2012, Abolhasani:2013zya, Lyth:2013sha}.
It has been shown that large non-Gaussianities with non-trivial shapes are generated. Considering our motivation in using
non-Gaussianity fingerprints in  LSS as a probe of inflationary universe, we would like to study the effects of large scale-dependent and orientation-dependent bispectrum on bias in these models.

The rest of the paper is organized as follows: In Section \ref{anisotropy-inflation} we present the anisotropic inflation model and the corresponding bispectrum which is used in subsequent analysis. In  Section \ref{SecBias} we review the basics of halo bias with non-Gaussian initial conditions. In Section \ref{general}, we present a general mathematical formulation for the anisotropic bispectrum in terms of spherical harmonics which can be used to calculate the halo bias in models with generic anisotropic bispectrum. In Section \ref{bias-analysis} we present our results of halo bias in the models of anisotropic inflation with the anisotropic bispectrum obtained  in  Section \ref{anisotropy-inflation}. We leave some technical issues of halo bias analysis into the Appendix.

In this paper we work with the natural unit in which  $c=\hbar=1$.


\section{Anisotropic Inflationary model}
\label{anisotropy-inflation}

In this section we review the anisotropic inflation model and its anisotropic bispectrum which will be used in subsequent analysis.

The best method to introduce statistical anisotropies is to incorporate $U(1)$ gauge fields or vector fields in models of inflation. However, due to conformal invariance of the standard Maxwell theory in an expanding background, the background gauge field and its quantum fluctuations are diluted during inflation. Therefore, in order to produce an almost scale-invariant power spectrum of gauge field fluctuations, one has to consider a time-dependent gauge kinetic coupling. This prescription was originally used in \cite{Turner:1987bw, Ratra:1991bn} in the context of primordial magnetic field.   An interesting model of anisotropic inflation is introduced in \cite{Watanabe:2009ct} in which it is shown that, with a suitably chosen gauge kinetic coupling,   the inflationary system  has an attractor solution in which the gauge field energy density reaches a small but observationally detectable fraction of the total energy density. As a result, the anisotropy produced are at the order of slow-roll parameters.

The Lagrangian of the system is
\ba
\label{action}
S= \int d^4 x \sqrt{-g} \left [ \frac{M_P^2}{2}
R -  \frac{1}{2} \partial_\mu \phi\partial^\mu \phi - \frac{f^2(\phi)}{4}F_{\mu \nu} F^{\mu \nu}  - V(\phi) \right] \, ,
\ea
in which $\phi$ is the inflaton field and $F_{\mu\nu} = \partial_\mu A_\nu - \partial_\nu A_\mu $ is the field strength associated with the $U(1)$ gauge field $A_\mu$.

The background is in the form of Bianchi I universe with the metric
\ba
ds^2 &=& - dt^2 + e^{2\alpha(t)}\left( e^{-4\sigma(t)}d x^2 +e^{2\sigma(t)}(d y^2 +d z^2) \right)  \nonumber\\
&=&   - dt^2 + a(t)^2 dx^2 + b(t)^2 (dy^2 + dz^2) \, .
\ea
Here $H \equiv \dot \alpha$ is interpreted as the average Hubble expansion rate, $H_a \equiv \dot a/a$
and $H_b \equiv \dot b/b$ are the expansion rates along the spatial directions $x$ and $y$
and $\dot \sigma/H \equiv (H_b - H_a)/ H$ is a measure of anisotropic expansion. We note that this metric enjoys only a two-dimensional rotational symmetry in $y-z$ plane.

The details of the dynamics of the system are given in \cite{Watanabe:2009ct, Abolhasani:2013zya}.
For the simple chaotic potential $V= m^2 \phi^2/2$, the conformal coupling (the time-dependent gauge kinetic coupling) is chosen to be
\ba
f(\phi) = \exp \left(   \frac{c \phi^2}{2 M_P^2}
\right) \, ,
\ea
where $c$ is a constant.   With $c\simeq 1$, one can check that the system admits the attractor solution in which the ratio of the gauge field energy density in the form of electric field energy density is a small and constant fraction of the total energy density. Defining the fraction of electric field energy density to the potential energy as
\ba
\label{R-def}
R \equiv \frac{\dot A^2 f(\phi)^2 e^{-2 \alpha}}{2 V} \, ,
\ea
during the attractor regime one obtains
\ba
R= \frac{c-1}{2 c} \epsilon_H = \frac{I}{2} \epsilon_H \, ,
\ea
in which $\epsilon_H \equiv -\dot H/H^2$ is the slow-roll parameter and
$I\equiv \frac{c-1}{c}$.

The power spectrum of the curvature perturbation is defined via
\ba
\langle \zeta_{\mathbf{k}}  \zeta_{\mathbf{k}'} \rangle  = (2 \pi)^3  P_\zeta(\mathbf{k})
\delta^3 ( \mathbf{k} + \mathbf{k}' )   \quad , \quad
{\cal P}_\zeta  \equiv \frac{k^3}{2 \pi^2} P_\zeta(\mathbf{k}) \, .
\ea
For the particular anisotropic inflation model  described above the power spectrum
was calculated in \cite{Abolhasani:2013zya, Dulaney:2010sq,  Gumrukcuoglu:2010yc, Watanabe:2010fh, Yamamoto:2012sq, Funakoshi:2012ym, Bartolo:2012sd, Emami:2013bk, Shiraishi2012,
Lyth:2013sha} which has the form
\begin{equation}
\label{power-eq0}
P_{\zeta}(\vec{k})=P_{0}\left(1+g_*(\hat{k}.\hat{n})^2\right) \, ,
\end{equation}
where the anisotropy parameter $g_*$ is given by
\begin{equation}
\label{g*-eq}
g_* \equiv -24I N(k_1)N(k_2) \, .
\end{equation}
Here $N(k_i)$ represents the number of e-folding that the mode of interest $k$ leaves the horizon. In our notation, the number of e-folding is counted backwards from the time of the end of inflation  by $a(N) = a_f \exp(N)$ so $N\leq 0$.  For example, $N= N_{CMB}=-60$ to solve the flatness and the horizon problem. As a result, $N(k)$ is calculated to be
\ba
\label{Nk}
N(k) - N_{CMB} = \ln \left(  \frac{k}{k_{CMB}} \right)  \, ,
\ea
in which $k_{CMB}$ represents the comoving mode which leaves the horizon at $N_{CMB}=-60$ e-folds before the end of inflation.  To satisfy the observational constraints from CMB and large scale structure we require $|g_*| <0.3$   \cite{Ade:2013ydc, Groeneboom:2009cb, Pullen:2010zy} corresponding to $I \lesssim 10^{-5}$ \cite{Dulaney:2010sq,  Gumrukcuoglu:2010yc, Watanabe:2010fh,  Emami:2013bk}.

The bispectrum of curvature perturbations, $B_{\zeta}(\vec k_{1}, \vec k_{2}, \vec k_{3})$,
is defined via
\ba
\label{bi- def}
\langle \zeta (\vec k_{1}) \zeta (\vec k_{2}) \zeta (\vec k_{3}) \rangle &\equiv& \left( 2 \pi \right)^3 \delta^3 \left( \vec k_{1} +  \vec k_{2} +  \vec k_{3}\right) B_{\zeta}(\vec k_{1}, \vec k_{2}, \vec k_{3})  \, .
\ea
The Bispectrum for the model of anisotropic inflation was calculated using in-in formalism in
\cite{Bartolo:2012sd, Shiraishi2012} and using $\delta N$ formalism in \cite{Abolhasani:2013zya} with the result
\ba
\label{Eq-bi}
B_\zeta(\vec k_1, \vec k_2, \vec k_3)=  288 I N(k_1) N(k_2) N(k_3) \left( C(\vec k_{1}, \vec k_{2}) P_0(k_{1})P_0(k_{2})
+ 2 \mathrm{perm.} \right)  \, .
\ea
Here the  anisotropic shape function $C(\vec{k}_{1},\vec{k}_{2})$ is defined as:
\ba
C(\vec k_{1}, \vec k_{2})\equiv\bigg{(}1 -   (\widehat k_1.\widehat n)^2  -   (\widehat k_2.\widehat n)^2 +
(\widehat k_1.\widehat n) \,  (\widehat k_2.\widehat n) \,  (\widehat k_1.\widehat k_2)  \bigg{)} \, ,
\ea
where $\hat{n}$ is the specific anisotropic direction in the sky. Note that in Eq. (\ref{Eq-bi}), $P_0(k_i)$ represents the isotropic power spectrum so all anisotropies are encoded in shape function
$C(\vec k_{1}, \vec k_{2})$ (and the appropriate permutations)
with the amplitude $288 I N(k_1) N(k_2) N(k_3)$.

It is instructive to look into the bispectrum in the squeezed limit in which one mode is much longer than the other two, say $k_3 \ll k_1 \simeq k_2$, so from the condition $\sum_i \vec k_i =0$ we also conclude that $\vec k_1 \simeq  - \vec k_2$.  In this limit, one obtains
\begin{equation}\label{bispectrum-ani}
B_\zeta(k_1,k_2,k_3)\simeq 24P_{0}(k_1)P_{0}(k_3)|g_*(k_1)|N(k_3)\times \left[1-(\hat{k}_{1}.\hat{n})^2-(\hat{k}_{3}.\hat{n})^2+(\hat{k}_{1}.\hat{n})(\hat{k}_{3}.\hat{n})(\hat{k}_{1}.\hat{k}_{3})\right]    \quad (k_3\ll k_1 \simeq k_2)
\end{equation}
in which to obtain the above result, Eq. (\ref{g*-eq}) has been used to express the parameter $I N(k_1) N(k_2)$ in terms of $g_*$.

The non-Gaussianity parameter $f_{NL}$ is defined in the squeezed limit $k_3 \ll k_1 \simeq k_2$ via \cite{Chen:2010xka, Komatsu:2010hc}
\ba
f_{NL} (\vec k_1, \vec k_2, \vec k_3) = \lim_{k_3 \rightarrow 0} \frac{5}{12}
\frac{B_\zeta(\vec k_1, \vec k_2, \vec k_3)}{P_\zeta(k_1) P_\zeta(k_3)} \, .
\ea
In general, $f_{NL}$ is an orientation-dependent and scale-dependent quantity. As an order of magnitude estimation,
and neglecting the logarithmic scale-dependence in $N{(k_i)}$,
we can define an orientation-dependent effective $f_{NL}^{eff}$ via
\begin{equation}\label{fnleff}
f^{eff}_{NL}=240IN_{CMB}^3C(\vec k_1, \vec k_2),
\end{equation}
keeping in mind that $N(k)=N_{CMB}+ \ln({k}/{k_{CMB}})$. Setting $N_{CMB}=60$, we can easily get $f_{NL}^{eff}\sim 60$ with $g_* \sim 0.1$, compatible with observational constraints.

A very interesting observation  from Eq.(\ref{bispectrum-ani}) is that when the long wavelength mode $\vec k_{3}$ is in the direction of anisotropy, (i.e. $\vec{k}_3\parallel \hat{n}$), then
the term inside the big-bracket in Eq.(\ref{bispectrum-ani}) vanishes. Consequently, in this configuration, we do not expect to see the  NG effects in LSS. We discuss this feature in more details in Sec.\ref{SecBias}.

For the subsequent analysis  we adopt the coordinate system such as the anisotropic direction
$\hat n$ coincides  with the  $\hat z$ direction in the spherical coordinates so the other momentum vectors are described by $\hat{k}_{1}=(\theta_{1},\psi_{1})$ and
$\hat{k}_{2}=(\theta_{2},\psi_{2})$, where $\theta$ and $\psi$ are the polar and azimuthal angles in spherical coordinates, respectively. For the future reference, we also need the angles between two arbitrary unit vectors $\hat q_i  = (\theta_{q_i}, \psi_{q_i})$ defined via $\cos \gamma = \hat q_1. \hat q_2$, which is
\begin{equation}
\label{angle-between}
\cos \gamma =\sin(\theta_{q_1})\sin(\theta_{q_2})\cos(\psi_{q_1}-\psi_{q_2})+
\cos\theta_{q_1}\cos\theta_{q_2} \, .
\end{equation}

The power spectrum and the bi-spectrum presented in Eqs. (\ref{power-eq0}) and (\ref{Eq-bi})
are for the particular model of anisotropic inflation as studied in \cite{Bartolo:2012sd, Shiraishi2012, Abolhasani:2013zya}. For a generic anisotropic model the most general power spectrum can be written as \cite{Pullen:2007tu}
\begin{equation}
\label{Eq-psaniso0}
P(\vec{k}) = P_{0}(k)  \left[1+\sum_ {LM}g_{LM}(k)Y_{LM}(\hat{k})\right],
\end{equation}
where $P_0$ is the isotropic power spectrum, $Y_{LM}(\hat{k})$ (with $L\geq 2$) are spherical harmonics and $g_{LM}(k)$ quantify the departure from statistical isotropy as
a function of wavenumber $k$. Since each Fourier mode $\vec{k}$ is related to $-\vec{k}$, in the case of real $g_{LM}(k)$, the multipole moment $L$ must be  even, and in the limit of $k\rightarrow 0$ we recover the isotropic power spectrum $P_{0}(k)$. However, in the general case,  (real and imaginary $g_{LM}$), we have
\begin{equation}
g^*_{L M}=(-1)^L g _{L-M} \, .
\end{equation}
This condition is imposed by the fact that the matter power spectrum is a real quantity.

Comparing Eq. (\ref{Eq-psaniso0}) with Eq. (\ref{power-eq0})  for our particular anisotropic inflation model we have
$g_{2 0} \propto g_* $ while the rest of $g_{LM}$ are zero.
In Section \ref{general} we extend the general definition of Eq. (\ref{Eq-psaniso0}) for the power spectrum to the  bispectrum and look into its implications in halo bias analysis.

\section{Bias}
\label{SecBias}

In this section we review the concept of  bias, a parameter that shows the dependence of dark matter halo abundance to the background dark matter density perturbations. The reader who is familiar with these analysis can directly jump to the next Sections in which we present our results of halo bias for anisotropic primordial power spectrum and bispectrum. It is worth to mention that in this work we are not interested in galaxy bias, which is the weighted integral of the halo bias, corresponding to the mechanism of halo occupation distribution (HOD).

In order to find an expression for the bias parameter we follow the work by Scoccimarro et al \cite{Scoccimarro:2011pz}.  However, there are other studies which  use Excursion Set Theory (EST) to calculate the halo bias.  In Adshead et al \cite{Adshead:2012hs} the authors solved the more complicated problem of non-spherical halos for which the collapse threshold becomes scale-dependent. In D'Aloisio et al. \cite{D'Aloisio:2012hr} EST is extended to path integral approach  taking into account the non-Markovianity effects of random walks in EST.

The halo bias relates the halo abundance to  the dark matter over-density. In Excursion Set Theory (EST) \cite{Bond1991} it is defined as
\begin{equation}
b(k,z)=\frac{\delta_{h}}{\delta_{m}},
\end{equation}
where $\delta_h$ is the halo over-density and $\delta_m$ is the matter density perturbation.
The EST framework is a very useful tool to calculate the abundance of structures. It is based on the concept of threshold crossing when we go from larger scales to smaller scales with the exclusion of the cloud-in-cloud effect which is present in Press-Schechter formalism\cite{Sheth:1998xe}.
At large scales  EST is known to reproduce the  initial condition while in small scales  it determines the local bias parameter with linear and nonlinear terms \cite{Mo:1996cn,Scoccimarro:2000gm} which are in reasonably good agreements with numerical simulations \cite{Smith:2006ne,Manera:2009ak,Manera:2009zu}.
It is worth to mention that the halo bias is a function of redshift and scales. This
scale-dependance is introduced by applying the initial non-Gaussian condition.
According to Appendix \ref{app-PBS} the large scale halo bias
can be treated in peak-background splitting \cite{Cole1989}. The idea of splitting of the
density contrast to short and long wavelength can be translated to a similar
splitting of Bardeen potential (correspondingly the matter density contrast) due to Poisson equation which depends on cosmological parameters.
The matter density in PBS can be written as
\begin{equation}
\rho=\bar{\rho}(1+\delta _{s}+\delta _{l}) \, .
\end{equation}
 The number density of formed structures with mass $m$ can
be expressed as a function of small scale statistics, (i.e. smalls scale power spectrum $P_{s}(k)$) and the background long wave-length perturbation $\delta_{l}$ (i.e.
$n=n[\delta_{l}(\vec{x}),P_{s}(k_{s});m]$)  \cite{Slosar2008}.
The bias parameter in the context of peak-background splitting (PBS) is described by  the fact that the background large scale over-density changes the critical threshold of spherical collapse \cite{Gunn:1972sv}. Therefor, the criteria for collapse becomes
\begin{equation}
\delta_{s}>\delta_{c}-\delta_{l},
\end{equation}
where $\delta_{s}$ is the matter density contrast of the structure, (the subscript ``s" stands for the short wavelength);
$\delta_{l}$ is the background (long-wavelength) density contrast and $\delta_{c}\simeq1.68$ is the critical density contrast in spherical collapse formalism 
\cite{Cole1989,Bardeen:1985tr} (for a review of PBS see Appendix \ref{app-PBS}).
Now in  order to find the bias parameter we have to compare the dark matter halo abundance,
in  cases with and without the presence of long wavelength (background) over-density.
For this task we use the EST approach. In Appendix \ref{app-EST},
we review the concept of EST in more details and we will derive the bias parameter using the PBS in EST context.


The primordial potential in Fourier space  can be translated into the
late time potential as
\ba
\Phi(k,z)=\frac{9}{10}\Phi_{ini} T(k)D(z)(1+z) \, .
\ea
Here $\Phi_{ini}$ represents the initial Bardeen potential sourced by
the inflaton field quantum fluctuations which is related to the curvature perturbation in radiation dominated via $\Phi_{ini}=2/3 {\cal{R}}$, $T(k)$ is the transfer function and
$D(z)$ is the growth function normalized to scale factor at early times. An important point here  is that we use the usual formalism of  the isotropic linear perturbation theory when we calculate
the effects of anisotropic NG on LSS observables. This is reasonable to first order  because after inflation ends  we recover the isotropic FRW Universe at the background level. The anisotropies are inherited only in seed perturbations which show themselves only through power-spectrum and bispectrum.

Now  one can relate the initial non-Gaussian potential to $\delta_{l}$ via
Poisson equation in sub-horizon scale and linear-regime
\begin{equation}
\delta_{l}=M(k,z)\Phi,
\end{equation}
where
\begin{equation}
M(k,z)= \frac{3k^2T(k)D(z)}{5\Omega^{0}_{m}H^2_{0}} \, .
\end{equation}
Here, $\Omega_0$ and $H_0$ are the matter fraction energy density and Hubble parameter at present time, respectively.
 In order to calculate the halo-bias term, we should calculate the
 effect of large scale perturbation, $\delta_{l}$,  on the Probability
 Distribution Function (PDF) density fluctuations . This yields the following relation between the Lagrangian halo number density and the PDF of density fluctuations \cite{Scoccimarro:2011pz}
\begin{equation}
1+\delta^{L}_{h}=\frac{\partial_{m}\int_{\infty}^{\delta_{c}}\Pi(\delta_{s},\sigma^2
_{m},\delta_{c};\delta_{l},\sigma_{l}^2)d\delta_{s}}{\partial_{m}\int_{\infty}^{\delta_{c}}\Pi_{0}(\delta_{s},\sigma^2
_{m},\delta_{c})d\delta_{s}} \, ,
\end{equation}
where $\Pi(\delta_{s},\sigma^2_{m},\delta_{c};\delta_{l},\sigma_{l}^2)$ is the conditional  PDF of density fluctuations for  $\delta_s$  with corresponding variance $\sigma_m$, when there is a background perturbation of $\delta_l$ and variance $\sigma_l$. The notation used in the conditional PDF,  means that  the variance  $\sigma_m$  at large scales converges to the value $\sigma_m = \sigma_{l}$, in contrast to the  unconditional  PDF of density fluctuations $\Pi_0(\delta_{s},\sigma^2
_{m},\delta_{c})$ in which the variance vanishes ($\sigma_m \to 0$) at large scales.
It will be relevant to define a quantity that shows
the probability of first up-crossing in the  time interval between $\sigma^2_{m}$ and $\sigma^2_{m}+d\sigma^2_{m}$ in EST language as
\begin{equation}
{\cal{F}}_{0}(\delta_{c},\sigma^2_{m})\equiv -\frac{\partial}{\partial\sigma_m^2}\int_{-\infty}^{\delta_{c}}\Pi_{0}(\delta_{s},\sigma^2_{m},\delta_{c})d\delta_{s} \, .
\end{equation}

From the above formalism, we can see the effect of primordial NG on LSS. Assuming that the Bardeen Potential has a local-type NG we have
\begin{equation}
\Phi=\phi+f_{NL}\phi^2 \, ,
\end{equation}
where $\phi$ is the Gaussian field. Now we can use the splitting idea on  $\phi$ by
applying $\phi=\phi_{l}+\phi_{s}$, where $\phi_{l}$ is the long wavelength mode of potential and $\phi_s$ is short wavelength
corresponding to the scale of structure. In Appendix \ref{app-PBS} we discuss how the above non-linear form can be generalized to a model with arbitrary shape of non-Gaussianity, in which case, the non-linear term generalizes into a kernel.

Now in the presence of primordial NG, the modes are not independent and the conditional PDF of density fluctuations is modified
by the non-Gaussian long-wavelength mode. In this case the PDF of density fluctuations will be a function of $\phi_{l}$ through the variance
and  also higher order cumulants ($c_{p}\equiv\langle\delta_{s}^{p}\rangle _{c}$) as
\begin{equation}
\Pi(\delta_{s},\sigma^2_{m},\delta_{c};\delta_{l},0)\rightarrow \Pi[\delta_{s},\sigma ^2(\phi),c_{p}(\phi),\delta_{c};\delta_{l}(\phi),0] \, .
\end{equation}

The non-Gaussian initial conditions introduces a dependence on higher-order cumulants which does not exist in the Gaussian case. These higher order cumulants depend on the long wavelength mode.
Under the assumption that all these effects are small,  using the  EST formalism we can Taylor expand the conditional PDF of density fluctuations, $\Pi$, around unconditional one, $\Pi_0$. The EST formalism with a sharp $k$-space filter and with the assumption of Gaussian initial conditions leads to a Markovian random walk condition for the density contrast value when  changing the mass scale/radius in  each step. This means that in the case of Markovianity we neglect the environmental dependence in halo formation process. Recently Maggiore and Riotto  \cite{Maggiore:2009rx,Maggiore:2009rv,Maggiore:2009rw} showed how to extend the EST with the path-integral method to include the non-Markovian condition. Also there are many follow up works where this effect on non-Gaussian halo bias is studied \cite{Paranjape:2011ak,Paranjape:2011wa,Musso:2011ck,Musso:2012qk,
Musso:2012ch} (for more details, see appendix \ref{app-PBS}).
In this work we study the effects on linear bias from the  anisotropic primordial NG and include only the first derivative contribution in Taylor expansion, Eq.(\ref{Eq-halobiasEPS}). As it was shown in Scoccimarro et al. \cite{Scoccimarro:2011pz}, the bias parameter calculated in first order of $f_{NL}$ is not sensitive to the Markovianity/non Markovianity condition. In this work we concentrate on NG at the order of $ f_{NL}$. It is worth mentioning that in higher order NG, such as in trispectrum analysis  yielding the $g_{NL}$ parameter, non-Markovianity is induced which  results in to a new scale-dependence in bias. The analysis in \cite{Paranjape:2011ak,Paranjape:2011wa,Musso:2011ck,Musso:2012qk, Musso:2012ch}
show that the departure from Markovian condition in  bias parameter is more significant for low mass ranges.  As we showed in appendix \ref{app-PBS}, up to first order in $f_{NL}$,  only the first two terms in Taylor expansion of PDF density fluctuations   appear. These terms are  derivatives of PDF density fluctuations  with respect to the long wavelength mode $\delta_l$ and the variance $\sigma_l$.  The higher order terms, corresponding to derivatives with respect to $c_p$ ($p\ge 3$), contribute to ${\cal{O}}(f^2_{NL})$ and ${\cal{O}}(g_{NL})$ bias.

 As mentioned above, in this work we consider only NG at the order of  $f_{NL}$.
The $p=1$ contribution, the first term in Taylor expansion, Eq.(\ref{Pdif}), is the usual
scale-independent linear bias from Gaussian perturbations.
Keeping in mind $b\equiv\delta_{h}/\delta_l$, for the first order linear bias ($b_{1L}$) we have
\begin{equation}
\label{b1l}
p=1: ~~ b_{1L}=\frac{\partial_m \int(\partial\Pi/\partial\delta_l)_0}{\partial _m\int\Pi_0}=\left[\frac{\partial}{\partial\delta_{l}}\ln(\frac{dn(\delta_{l})}{d\ln{m}})\right],
\end{equation}
which can be written as:
\begin{equation}
b_{1L}=\frac{\partial}{\partial\delta_l}\ln(n(\delta_l)) \, .
\end{equation}

In the presence of primordial non-Gaussianity, there are new contributions from higher order cumulants $ p \ge 2$. As a result, the next to leading order term gives
\begin{equation}
p=2: ~~ b_{2L}=\frac{\partial_m [I_{21}\int \partial\Pi_0/\partial\sigma_m^2]}{M(k)\partial _m\int\Pi_0} \, ,
\end{equation}
which in general is a scale-dependent correction to the leading order, scale-independent bias, Eq. \eqref{b1l}.
The key quantity here is $I_{21}$ which is  the derivative of second cumulant $\sigma_m^2$, $(p=2)$, with respect to the
long wavelength mode $\phi_l$  which is obtained as \cite{Scoccimarro:2011pz}
\begin{equation}
\label{I21}
I_{21}(k,m)=\frac{1}{P_{\phi}(k)}\int B_{\hat{\delta}\hat{\delta}\phi}(q,k-q,-k)d^3q,
\end{equation}
where $B_{\hat{\delta}\hat{\delta}\phi}$ is the cross bispectrum of small-scale smoothed density $\hat{\delta}$
and $\phi$. As a result,  $I_{21}$ is the quantity which we are looking for in the case of non-Gaussian initial condition which introduces scale-dependent bias at the order of ${\cal O}(f_{NL})$.

So far only the Lagrangian bias appeared because
peaks are those of the initial density field (linearly extrapolated). Making the
standard assumptions that halos move coherently with the underlying dark matter,
and using the techniques outlined in \cite{Mo:1995cs,Catelan:1997qw,Efstathiou:1988tk,Cole:1989vx}, one can obtain the final Eulerian bias in linear order as
\begin{equation}\label{eq-Eubia}
b_{E}=1+b_{1L}+b_{2L};
\end{equation}
where the linear bias is
\begin{equation}\label{b1l}
b_{1L}=\frac{2}{\delta_c}\partial_{\ln\sigma_m^2}\ln(\sigma_m^2{\cal{F}})=b_{1L(G)}+b_{1L(NG)} \, .
\end{equation}
A very important point is that in above equation we have omitted the subscript of ${\cal{F}}$, which means that the non-Gaussianity changes the mass function of the structures so the first linear term will have a contribution from primordial non-Gaussianity. Consequently, we have
\begin{equation}
b_{1L(G)}=\frac{2}{\delta_c}\partial_{\ln\sigma_m^2}\ln(\sigma_m^2{\cal{F}}_{0})
\end{equation}
and
\begin{equation}
b_{1L(NG)}=\frac{2}{\delta_c}\partial_{\ln\sigma_m^2}\ln(\sigma_m^2{\cal{R}}_{NG})=\frac{\partial \ln {\cal{R}}_{NG}(m,f_{NL})}{\partial\delta_l},
\end{equation}
where ${\cal{F}}={\cal{R}}_{NG}{\cal{F}}_{0}$, and ${\cal{R}}_{NG}$  comes from the deviation of PDF density fluctuations  from the Gaussian case \cite{LoVerde:2007ri,Sefusatti:2006eu}. In other words, the effects of  non-Gaussianity appeared both in the  mass function and in the power spectrum via scale-dependent bias parameter. Since in this work we are interested in the scale-dependence features of bias, the contribution of ${\cal{R}}_{NG}$ is not much of interest.  For the Gaussian case we use the Sheth-Tormen \cite{Sheth:1999mn} Gaussian mass function. For the non-Gaussian mass function  effect we use the results of  \cite{Verde:2010wp} in which  the non-Gaussian mass function is expanded   in the Press-Schechter framework \cite{Press:1973iz} such that 
\begin{equation}
\label{RNG}
{\cal{R}}_{NG}(m,f_{NL})=1+\frac{1}{6}x(x^2-3)s_{3}(x)-\frac{1}{6}(x-1/x)\frac{ds_3(x)}{d\ln(x)},
\end{equation}
where $x\equiv{\delta_c}/{\sigma_M}$ and $\delta_c=1.68$ is the critical density and $s_{3}$ is the reduced skewness defined as
\begin{equation}
s_{3}(R)\equiv\frac{\langle\delta^3_R\rangle}{\langle\delta^2_R\rangle ^{3/2}}=\frac{\langle\delta^3_R\rangle}{\sigma^3_m} \, .
\end{equation}
The skewness is related to the matter bispectrum as
\begin{equation}
\langle\delta_R^3\rangle=\int \frac{d^3q_1}{(2\pi)^3}\frac{d^3q_2}{(2\pi)^3}W(Rq_1)W(Rq_2)W(Rq_{12})M(q_1,z)M(q_2,z)M(q_{12},z)B_0(q_1,q_2,q_{12}),
\end{equation}
where $\vec{q}_{12}=-(\vec{q}_{1}+\vec{q}_{2})$, $W$ is the window function in Fourier space and $R$ is the smoothing scale. For more details see Appendix \ref{app-EST}.

On the other hand, the scale-dependent bias can be rewritten as
\begin{equation}
b_{2L}=\frac{I_{21}(k,m)}{2\sigma_m^2 M(k,z)}\delta_c b_{1L}+\frac{1}{M(k,z)}\partial_{\ln\sigma_m^2}(\frac{I_{21}(k,m)}{\sigma_m^2}) \, .
\end{equation}
So the total Eulerian bias, up to first order in $f_{NL}$, can be split into the
 scale-independent term, $b_{si}$,  and scale-dependent term, $b_{sd}$,  as
\begin{equation}
b_{t(E)}=b_{si}+b_{sd}
\end{equation}
where $b_{si}$ and $b_{sd}$ are
\begin{equation}
b_{si}\equiv b_{G}+b_{1L(NG)},
\end{equation}
with $b_{G}\equiv 1+b_{1L(G)}$ and
\begin{equation}
b_{sd}=b_{2L}+{\cal{O}}(f^2_{NL})+{\cal{O}}(g_{NL})+...
\end{equation}

In  next section we find the scale-independent and the scale-dependent bias for general anisotropic initial power spectrum and discuss the effects of the primordial anisotropies on the bias parameter.

\section{General formulation of Anisotropic Bias}
\label{general}

In this section we calculate the bias parameter, assuming a general model independent primordial anisotropy. For this task we write the power spectrum and bispectrum in their most general anisotropic forms. Then we find the possible configurations that give
the scale-dependent bias.

As mentioned in section \ref{anisotropy-inflation}, the general anisotropic power spectrum can be written as  \cite{Pullen:2007tu}
\begin{equation}
\label{Eq-psaniso}
P(\vec{k}) = P_{0}(k)  \left[1+\sum_ {LM}g_{LM}(k)Y_{LM}(\hat{k})\right],
\end{equation}
which is a generalization of our anisotropic power spectrum defined in Eq.(\ref{power-eq0}), where all the $k$-dependence is buried in the coefficient $g_{LM}$.
In this work we assume that the mechanism of spherical collapse for structure formation is still applicable in our model with primordial anisotropic perturbations.  This is motivated from the fact that after inflation the background becomes isotropic and anisotropies are encoded only on primordial perturbations. Secondly, the collapse mechanism is a local process so up to first order it is not affected by large scale anisotropies. As a result, we assume that the transfer function at leading order is not affected by the primordial anisotropies. Having this said, it would be interesting to perform the analysis with the above assumptions being relaxed, but this  is beyond the scope of this work.

First of all the variance is defined through the linear matter power spectrum
\begin{equation}
\sigma^2(M,z)=\int\frac{d^3k}{(2\pi)^3}P_{L}(k)W^2(kR),
\end{equation}
where $P_{L}$ and $W(kR)$ are  the linear matter  power spectrum and window function in Fourier space which are discussed in details
in the Appendix \ref{app-EST}.
The anisotropic variance reads as
\begin{equation}\label{sigmaAA}
\sigma_A^2(m,z)=\int d\psi d(\cos\theta)k^2dkP_0(k,z)W(kR)\left[1+\sum_{LM}g_{LM}(k)Y_{LM}\right] \, .
\end{equation}
The spherical harmonics are orthonormal  via
\ba
\int Y_{lm} Y^*_{l'm'} d\Omega = \delta_{ll'}\delta_{mm'}  \, .
\ea
As a result, noting that $Y_{00}=1/\sqrt{4\pi}$, we have
\ba
\int Y_{lm} d\Omega = \sqrt{4\pi} \int Y_{lm} Y^*_{00}\, d\Omega =\sqrt{4\pi} \, \delta_{l0} \delta_{m0}  \, ,
\ea
so the variance, Eq.(\ref{sigmaAA}), can be simplified to
\begin{equation}
\label{var-gen}
\sigma_A^2(m,z)=4 \pi \int k^2dkP_0(k,z)W(kR)\left[1+g_{00}(k)/\sqrt{4\pi}\right] \, .
\end{equation}
Ignoring the scale-dependence of $g_{00}$, we have
\begin{equation}
\sigma_A^2(m,z) \simeq \sigma_0^2(m,z) \left[1+g_{00}/\sqrt{4\pi}\right],
\end{equation}
with
\begin{equation}
\sigma_0^2(m,z)=4 \pi \int k^2dkP_0(k,z)W(kR) \, .
\end{equation}
Similarly, we can generalize the bispectrum in squeezed limit, where $k_3 \ll k_1\approx k_2$, as follows
\begin{equation}
\label{bi-gen}
B_{\zeta}(\vec{k}_{1},\vec{k}_{2},\vec{k}_{3})\approx 2\sum_{L,l,m} c_{Llm}{\cal{P}}_{L}(\cos\theta_{3}){Y}_{l,m}(\theta_{1},\psi_1)\left(P_0(\vec{k}_1)P_0(\vec{k}_3)\right)
\end{equation}
where ${\cal{P}}_L$ is the Legendre function, ${Y}_{l,m}(\theta_{1},\phi_1)$ are the spherical harmonics, by which the whole direction-dependence of bispectrum is encoded and the $c_{Llm}$ are the $k$-dependent coefficients. This decomposition is possible because we set the $z$ coordinate along the anisotropy direction and rotate the $x-y$ plane in Fourier (k-space) such that the azimuthal angle of $k_3$ is set to zero. We can do this rotation because of the symmetry in the $x-y$ plane. This kind of decomposition, i.e. decomposing into
a $k$-dependent sector ($c_{Llm}$) and the angular-dependent part is especially useful to find the corresponding $I_{21}$ term in the case of  generic anisotropic bispectrum and the corresponding scale-dependent bias.

Since in the real space the bispectrum is a real quantity, and noting that the bispectrum above is defined in Fourier space,  the following condition holds
\begin{equation}\label{cllm}
c^*_{Llm}=(-1)^{L+l} c_{Ll -m} \, .
\end{equation}
 Now, using  Eq.(\ref{bi-gen}), we can write $I_{21}$ as
 \begin{equation}\label{I21general}
 I_{21}=2\sum _{L,l,m}\int d\psi_1 d(\cos\theta_1)q^2dq M_{m,z}(q)M_{m,z}(|k-q|) c_{Llm}(q,k){\cal{P}}_L(\theta_3)Y_{l,m}(\theta_1,\psi_1)P_0(q),
 \end{equation}
which, again using the orthonormality of spherical harmonics, simplifies to
 \begin{equation}\label{I21general}
 I_{21}({\bf k},m,z)=4 \sqrt{\pi}\sum _{L} {\cal{P}}_L(\cos \theta_k)\times\int q^2dq M_{m,z}(q)M_{m,z}(|k-q|) c_{L,l=0,m=0}(q,k)P_0(q) \, .
 \end{equation}

This is an interesting result showing that the anisotropic bispectrum induces an anisotropic, scale-dependent bias through the Legendre function of the angle between the anisotropic direction $\hat{n}$ and the long wavelength mode. It is also interesting to note that there is a {\it{Selection Rule}} for scale-dependent bias at ${\cal{O}}(f_{NL})$, where the harmonic numbers  of
the short wavelength must vanish ($l=0$, $m=0$). All the other parts of anisotropic bispectrum with $l,m \neq 0$ do not contribute to the leading order scale-dependent bias.

It is worth to mention that the reality assumption of the bispectrum implies
\begin{equation}
c^*_{L00}=(-1)^Lc_{L00},
\end{equation}
where in the case of real $c_{L00}$,  $L$ is even.  Now, if we define a
direction-independent  parameter $I^{di}_{(L)21}$ as
 \begin{equation}\label{I21general}
 I^{di}_{(L)21}(k,m,z)\equiv4 \sqrt{\pi}\times\int q^2dq M_{m,z}(q)M_{m,z}(|k-q|) c_{L00}(q,k)P_0(q),
 \end{equation}
 then $I_{21}$ will be re-expressed as
 \begin{equation}
 I_{21}(k,m,z,\theta_k)=\sum _{L} {\cal{P}}_L(\cos \theta_k) I^{di}_{(L)21}(k,m,z)
 \end{equation}
where $\theta_k$ is the angle between $\hat{n}$ and $\vec{k}$.
Now the scale-dependent bias becomes
\begin{equation}
\label{Eq.b2l-b}
b_{2L}(k,z,m,\theta_k)=\frac{\sum _{L} {\cal{P}}_L(\cos \theta_k) I^{di}_{(L)21}(k,m,z)}{2\sigma_A^2(m,z) M(k,z)}\delta_c b_{1L}+\frac{1}{M(k,z)}\partial_{\ln\sigma_A^2}(\frac{\sum _{L} {\cal{P}}_L(\cos \gamma) I^{di}_{(L)21}(k,m,z)}{\sigma_{A}^2(m,z)})
\end{equation}
where in this case $\sigma_A^2$ is defined as in Eq.(\ref{var-gen}).

In next Section, as a specific example of the general formulation presented above, we study our model of anisotropic inflation introduced in Sec. (\ref{anisotropy-inflation}), corresponding to   $L=2, l=0, m=0$.


\section{Anisotropic Bias for gauge field inflationary models}
\label{bias-analysis}

In this Section we study the LSS bias in the model of  anisotropic inflation  as a special example of general formulation developed in previous Section. For a related work with a phenomenological modeling of bispectrum and its implication for bias see \cite{Shiraishi:2013sv}.
As in the previous Section, we continue with  our simplifying  assumptions of spherical collapse and take the transfer function to be that of the isotropic background.  We investigate the change in PDF of density perturbations and the corresponding cumulants, and the effect of these changes on bias.

In our model the variance is modified due to the fact that we use anisotropic power spectrum. However, we show below that it is not direction-dependent (as it was shown in general case).  Without loss of generality, we can assume that the anisotropy is pointed along the  $z$-direction in spherical coordinates ($\hat n = \hat z$). Then, starting from Eq. (\ref{power-eq0}) for the primordial anisotropic power spectrum, we have
\begin{equation}\label{sigmaA}
\sigma^2_{A}(m,z)=\int d\psi d\cos\theta dk k^2P_0(k,z)[1+g_*(k)\cos^2\theta]W(kR)\simeq(1+ \frac{g_*}{3})\sigma^2_m \, ,
\end{equation}
where $\theta$ and $\psi$ are spherical coordinate angles and
$g_*(k) = - 24 I N(k_1) N (k_2)$. Note that $\sigma^2_A$ is the variance obtained from the full anisotropic power spectrum, where $\sigma_m^2$ is the variance corresponding to the  isotropic part. Since the scale-dependence of $g_*$ is logarithmic through $N(k_i)$, as given in Eq. (\ref{Nk}), as a first approximation we can ignore its scale-dependence so we have the last approximate equality in Eq. (\ref{sigmaA}). Note that, as we mentioned before, the variance does not have any direction-dependence, which is somewhat an obvious observation, since one should integrate over the full 3D Fourier space to obtain the variance, eliminating any
direction present in power spectrum. However, it is interesting to note that depending on the sign of anisotropy parameter $g_*$ the correction to the variance due to anisotropy enhances or suppresses the leading order term. From the above variance, we can obtain the leading order scale-independent bias $b_{1L}$.


\begin{figure}[t]
\includegraphics[scale=1.3]{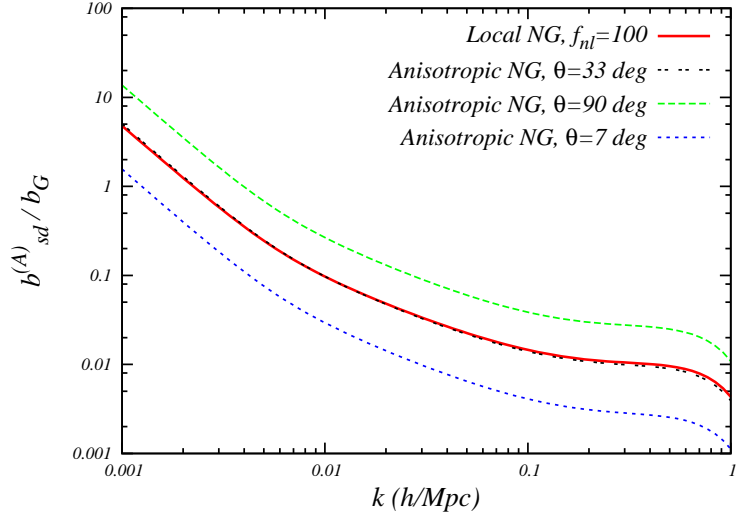}
\caption {The fraction of scale-dependent anisotropic bias Eq. \eqref{Eq.b2l}, to the Gaussian bias, $b^{(A)}_{sd}/b_{G}$, is plotted versus the wavenumber for different angles between the long wavelength mode and the anisotropy direction. The black double-dashed curve, the green long-dashed curve and the blue dotted curve, respectively, are for
$\theta_{k} = 33^{\circ}, \theta_{k}=90^{\circ}$ and  $\theta_{k}=7^{\circ}$. The solid red curve indicates the local non-Gaussianity with $f_{NL}=100$. For the anisotropic bias, we
choose $I \simeq 1.9 \times 10^{-6}$ such that $f^{eff}_{NL}=100$ from Eq. (\ref{fnleff}). The redshift is $z=0$.} \label{figtheta}
\end{figure}

  \begin{figure}[!ht]
    \begin{minipage}{0.49\linewidth}
\includegraphics[scale=1.0]{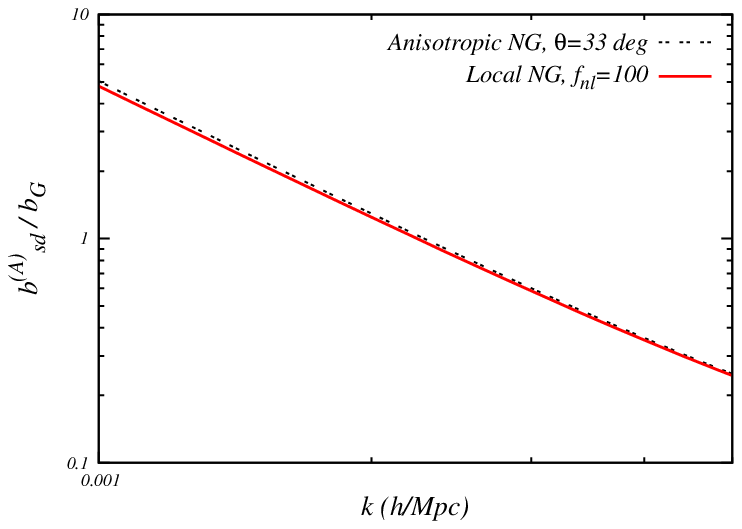}
      \caption{This figure is a zoom-in of Fig. \ref{figtheta} for long wavelength modes where we compare $b^{(A)}_{sd}/b_{G}$ in our model with $\theta_{k} =33^{\circ}$ to the local non-Gaussian model with $f_{NL}=100$. This shows small deviation between the two models due to a mild scale-dependence of $g_*$ parameter.}\label{fig:left}
    \end{minipage}
    \hfill
    \begin{minipage}{0.49\linewidth}
\includegraphics[scale=1.0]{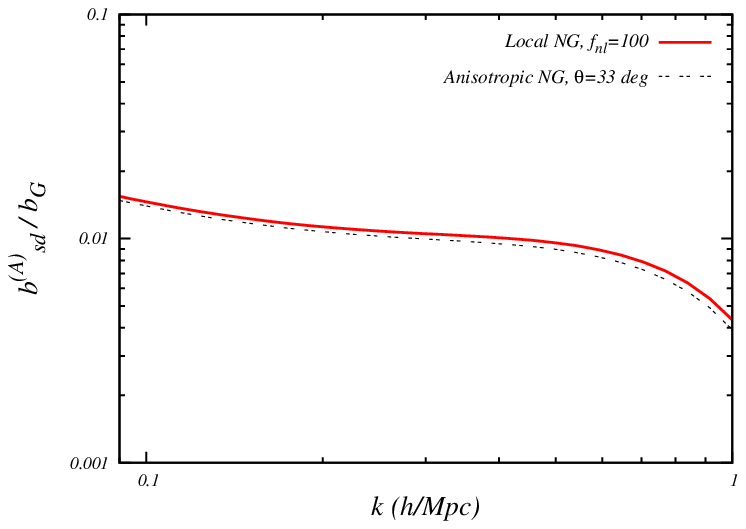}
      \caption{This figure is a zoom-in of Fig. \ref{figtheta} for short wavelength modes where again we compare $b^{(A)}_{sd}/b_{G}$ in our model for $\theta_{k} =33^{\circ}$ to the local non-Gaussian model with $f_{NL}=100$.}\label{fig:right}
    \end{minipage}
     \begin{minipage}{0.49\linewidth}
\includegraphics[scale=1.0]{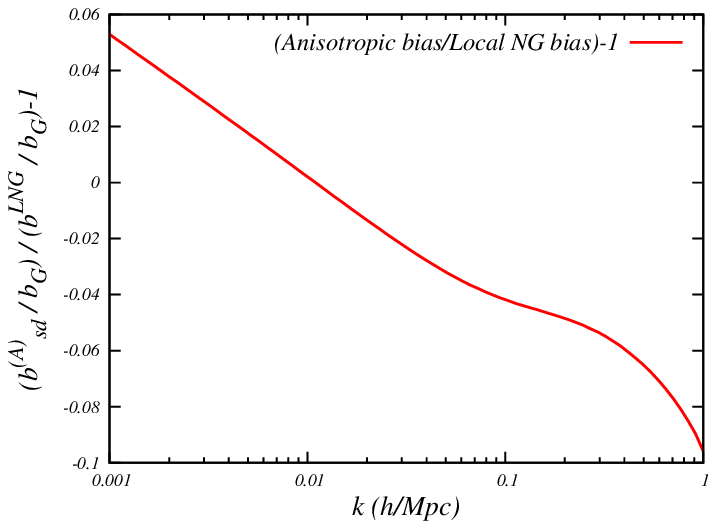}
      \caption{ This figure shows the relative magnitude of  the scale-dependent bias parameter in our  anisotropic model with $\theta_{k} =33^{\circ}$ compared to the local non-Gaussian model with  $f_{NL}=100$ }\label{fig:down}
    \end{minipage}

  \end{figure}

\begin{figure}[t]
\includegraphics[scale=1.3]{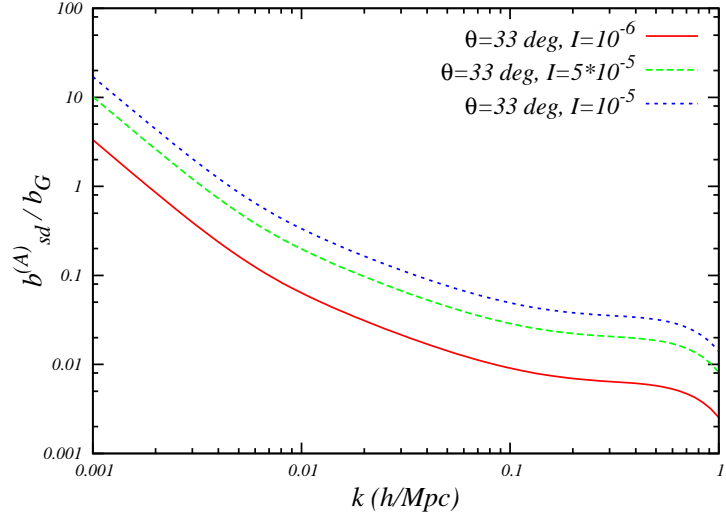}
\caption {$b^{(A)}_{sd}/b_{G}$ versus wavenumber is plotted  with  $\theta_{k} =33^{\circ}$ for three different values of $I$. The solid red curve,  the green dashed curve
and blue dotted curve, respectively, are for $I=10^{-6}, I=5\times10^{-6}$ and $I=10^{-5}$.
In all cases we set
the redshift to z=0.} \label{biasI}
\end{figure}

\begin{figure}[t]
\includegraphics[scale=1.3]{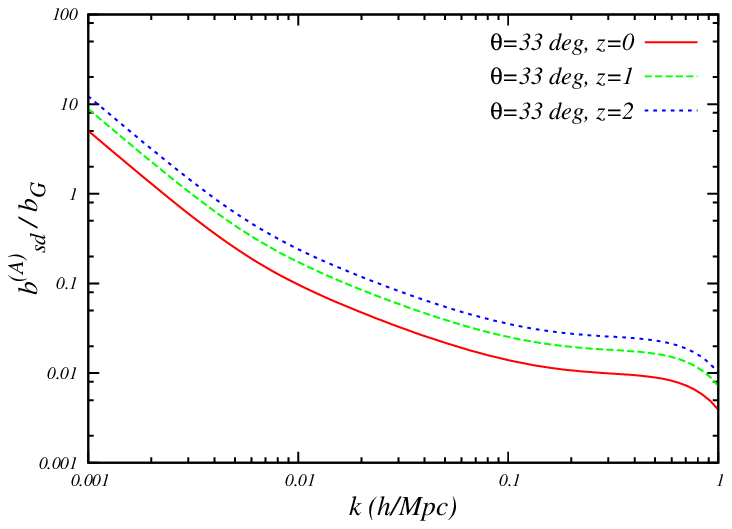}
\caption {$b^{(A)}_{sd}/b_{G}$ versus wavenumber is plotted  with $I=10^{-6}$ and
$\theta_{k} =33^{\circ}$ for three different redshifts. The solid red curve, the green dashed curve and the dotted blue curve, respectively, are for $z=0, z=1$ and $z=2$.} \label{biasz}
\end{figure}

As for the next step, we obtain the bias from the anisotropic bispectrum which is both
scale-dependent and direction-dependent. This would be the main effect of anisotropy in the bias parameter. In order to obtain $b_{2L}$, we need $I_{21}$ where the information from primordial bispectrum is encoded. We use the squeezed limit ($k_{3}\ll k_2\simeq k_1$) bispectrum predicted by the model,  Eq.\eqref{bispectrum-ani}. Because of the symmetry in $x-y$ plane we rotate the long wavelength mode such that $\psi_{\hat k_3}=0$. Inserting the bispectrum to $I_{21}$,
Eq.\eqref{I21} results in
\begin{eqnarray}
I_{21}({\bf k},z,m)&=&24\int d\psi_q d(\cos\theta_q)dq q^2 M_{m}(q,z)M_{m}(|k-q|,z)P_\phi(q) N(k)|g_*(q)| \\ \nonumber &\times &\left[1-\cos^2\theta_q-\cos^2\theta_k+\cos\theta_q\cos\theta _k\left(\sin\theta_q\sin\theta_k\cos\psi_q+\cos\theta_q\cos\theta_k\right)\right] \, ,
\end{eqnarray}
where $k$ and $q$ correspond to long and short wavelength ($k_3$ and $k_2$ in
Eq. \eqref{bispectrum-ani}), and $\theta$, $\psi$ are polar and azimuthal angles in spherical coordinates, defined by the angle between the anisotropy direction $\hat n$ and wavenumbers $\hat{q}$ and $\hat{k}$ respectively
\begin{equation}
\label{angle-between1}
\cos \gamma_q = \hat{q}.\hat{n}=\sin\theta_{\hat{n}}\sin\theta_{\hat{q}}\cos(\psi_{\hat{n}}-\psi_{\hat{q}})+
\cos\theta_{\hat{n}}\cos\theta_{\hat{q}},
\end{equation}
\begin{equation}
\label{angle-between2}
\cos \gamma_k =\hat{k}.\hat{n}=\sin\theta_{\hat{n}}\sin\theta_{\hat{k}}\cos(\psi_{\hat{n}}-\psi_{\hat{k}})+
\cos\theta_{\hat{n}}\cos\theta_{\hat{k}} \, .
\end{equation}
Now we can integrate the angular dependence $\gamma_q$ which yields
\begin{equation}\label{I21A}
I_{21}(k,z,m,\theta_k)=64\pi N(k)(1-\cos^2\theta_{k})\int dq q^2 M_{m,z}(q)M_{m,z}(|k-q|)P_\phi(q) |g_*(q)| \, .
\end{equation}
It is interesting to note that the orientation-dependence appears in the form of
$\sin^2\theta_{k}$. As a result, the  bias vanishes when $\sin \theta_{k}=0$, i.e. when the long wavelength mode is aligned with the anisotropic direction. This result  originates from the fact that in this specific direction the bispectrum vanishes in squeezed limit as one can check from Eq. \eqref{bispectrum-ani}.
Furthermore, since in the model under consideration  $g_*$ has a mild scale-dependence via logarithmic correction in  $N(k)$,  we observe that there is an extra but mild k-dependent factor in $I_{21}$ in comparison with the standard local non-Gaussian shape \cite{Scoccimarro:2011pz,Sefusatti:2012ye}.  Besides that,  $I_{21}$ linearly depends on $I$ which is
the free parameter of  the anisotropic inflationary model.
Now, by using the variance and $I_{21} $ parameter obtained above we can find the bias parameter in the anisotropic model by
\begin{equation}\label{Eq.b1l}
b_{1L}=\frac{\partial_m \int(\partial\Pi/\partial\delta_l)_0}{\partial _m\int\Pi_0}=\left[\frac{\partial}{\partial\delta_{l}}\ln(\frac{dn(\delta_{l})}{d\ln m})\right]=b_{1L(G)}+b_{1L(NG)},
\end{equation}
where, for the first order linear bias in the case of anisotropy, we have
\begin{equation}
b_{(A)1L(NG)}=\frac{2}{\delta_c}\partial_{\ln\sigma_A^2}\ln(\sigma_A^2{\cal{R}}_{(A)NG})=\frac{\partial \ln {\cal{R}}_{(A)NG}(m,f_{NL})}{\partial\delta_l},
\end{equation}
in which the subscript $(A)$ has been added to point out that the parameters are obtained in the presence of anisotropy. Here ${\cal{R}}_{(A)NG}$ is the anisotropic non-Gaussian correction to the  PDF of density fluctuations  defined by
\begin{equation}
{\cal{R}}_{(A)NG}(m,f_{NL})=1+\frac{1}{6}x_A(x_A^2-3)s_{(A)3}(x_A)-\frac{1}{6}(x_A-1/x_A)\frac{ds_3(x_A)}{d\ln(x_A)},
\end{equation}
where $x_A\equiv{\delta_c}/{\sigma_A}$ and $\delta_c=1.68$ is the critical density and  $s_{(A)3}$ is the reduced skewness defined as
\begin{equation}
s_{(A)3}(R)\equiv\frac{\langle\delta^3_R\rangle}{\langle\delta^2_{(A)R}\rangle ^{3/2}}=\frac{\langle\delta^3_{(A)R}\rangle}{\sigma^3_A} \, .
\end{equation}
Note that, in the above formula, we have to use the full variance $\sigma_A$ including the correction due to anisotropy. The anisotropic skewness is also defined as
\begin{equation}
\langle\delta^3_{(A)R}\rangle=\int \frac{d^3q_1}{(2\pi)^3}\frac{d^3q_2}{(2\pi)^3}W(Rq_1)W(Rq_2)W(Rq_{12})M(q_1,z)M(q_2,z)M(q_{12},z)B_A(q_1,q_2,q_{12}),
\end{equation}
where $B_{A}$ is the anisotropic bispectrum.

The next order term in bias will give the scale-dependent and direction-dependent effect
\begin{equation}
\label{Eq.b2l}
b^{(A)}_{sd}=b_{(A)2L}(k,z,m,\theta_k)=\frac{I_{21}(k,z,m,\theta)}{2\sigma_A(m,z) ^2 M(k,z)}\delta_c b_{1L}+\frac{1}{M(k,z)}\partial_{\ln\sigma_A^2}(\frac{I_{21}(k,m,\theta_k)}{\sigma_A^2(m,z)}) \, .
\end{equation}
So the total Eulerian anisotropic bias is defined as
\begin{equation}
b_{t(E)}=1+b_{1L(G)}+b_{(A)1L(NG)}+b_{(A)2L} \, .
\end{equation}
The first three terms above gives the scale-independent bias, whereas the last term is scale-dependent and direction-dependent bias due to the anisotropic bispectrum. Since the model we consider has a bispectrum shape very close to the standard local non-Gaussian shape and since $b_{1L(NG)}$ is small in comparison with $b_{sd}$ in local shape, \cite{Sefusatti:2012ye}, we can ignore  $b_{1L(NG)}$ term in our analysis.

In Fig. \ref{figtheta}  we plot the relative magnitude of the scale-dependent bias
Eq. \eqref{Eq.b2l} to Gaussian bias, $b^{(A)}_{(sd)}/b_{G}$, versus the wavenumber.
As expected we have approximately  $k^{-2}$ scale-dependence similar to the local non-Gaussian shape. However, this $k$-dependence is slightly different because of mild dependence of $g_{*}$ to wavenumber. A completely new feature is the direction-dependence of bias originated form primordial anisotropy, proportional to $\sin^2\theta_{k_3}$. In the case of $\theta_{k_3}=\pi/2$ we have the maximum scale-dependent bias, whereas at angles $\theta_{k_3} =0, \pi$, as the anisotropic bispectrum vanishes, the scale-dependent bias also vanishes accordingly.

In order to compare our results with the conventional local non-Gaussian models, we set $\theta_{k_3}=33^{\circ}$ to have $f^{eff}_{NL}\simeq 100$ and then compare the bias in this specific direction with the bias in local shape with amplitude $f_{NL}=100$.  In Fig \ref{fig:left}  and Fig. \ref{fig:right}, for both long and short wavelengths, we compare the fraction of scale-dependent  bias, $b_{(sd)}/b_{G}$, from
local non-Gaussianity compared to our anisotropic model.  In Fig. \ref{fig:down} we compare  their relative magnitudes. Since $g^{1/2}_* \propto N(k)=N_{CMB}+\ln(k/k_{CMB})$ the bias of our model is slightly higher than local non-Gaussian case for long wavelength modes
while it is slightly lower for short wavelength modes.

The free parameter of our model is $I$. In Fig. \ref{biasI}  we plot $b^{(A)}_{(sd)}/b_{G}$ versus wave number for different values of $I$. The $I$-dependence of bias is linear according to Eq. (\ref{I21A}). In Fig. \ref{biasz}   we plot the redshift-dependence of $b^{(A)}_{(sd)}/b_{G}$. As can be seen, at higher redshifts we have higher bias. This is the standard expectation, since at earlier times the non-linearity of local, small scale perturbations was weaker and thus they are more sensitive to large scale perturbations, resulting in larger bias.

\section{Conclusion and Discussions}
\label{discussion}

The large scale structure observations, like the statistics of rare objects and the
scale-dependence of bias parameter,
can be used as complementary cosmological observations to CMB data to constrain the inflationary models. In this work we obtained the scale-dependent and the direction-dependent bias of dark matter halos in anisotropic inflationary models.

The anisotropic model studied in this work  has a bispectrum shape very close to the local non-Gaussian shape with an extra  mild $k$-dependence and also a direction-dependence. We showed that the bias parameter is mainly influenced by the angle ($\theta_{k_3}$) between the anisotropy direction and the long wavelength mode $\vec k_3$ in the squeezed limit by a factor of $\sin^2\theta_{k_3}$. An interesting observation is that
the scale-dependent bias vanishes at linear order  along the direction $\theta_{k_3}=0, \pi$. This can be explained by the fact that the bispectrum vanishes in squeezed limit when the long wavelength mode is aligned with the anisotropy direction. This means that, at the level of bispectrum, the model reduces to the Gaussian model along this direction.

As it is clear from the PBS formalism, the bias parameter is  mainly affected by the long wavelength mode. We can see this explicitly in our work where the angle between the anisotropy direction and the long mode appears in the bias formula.

The free parameter of our model is $I$. Constraints from the CMB and LSS observations require  $I \lesssim 10^{-5}$ \cite{Dulaney:2010sq,  Gumrukcuoglu:2010yc, Watanabe:2010fh,  Emami:2013bk}. Roughly speaking, our bias parameter analysis also
imply the same order of magnitude for $I$. Considering the fact that the observations are now in good agreement with a Gaussian bias we expect that  in quasi-linear regime (where we have the strict constraint on bias) the ratio  $b^{(A)}_{sd}/b_{G}$ is at the order of one so
from our results we also find  $I \lesssim 10^{-5}$.

In this work we also formulated the general, model-independent anisotropic scale-dependent bias at linear order. For this purpose, by assuming the rotational symmetry in $x-y$ plane of Fourier space, we modeled the bispectrum as a function of spherical harmonics, $Y_{lm}$, and the  Legendre functions, ${\cal P}_{L}$, of the angles between the short/long wavelength modes and the anisotropy direction. Interestingly, we find a selection rule for scale-dependent anisotropic bias, which shows that the bias parameter only responses to the $Y_{00}$ part of the anisotropic bispectrum. We also show that  $b_{sd}$ is independent of short wavelength mode and its direction. This is an obvious check for the formalism, since the bias should be only a function of the long wavelength mode.

In this work we assumed that the  mechanism of spherical collapse and the conventional form of the transfer function are still applicable. This is motivated from the fact that the collapse of structures is a local mechanism which is  less affected by the large scale anisotropies. Having this said in principle it is an interesting question to see how one can generalize the spherical collapse mechanism in
the presence of primordial  anisotropic fluctuations. On the other hand,  the modification of the transfer function due to Boltzmann and perturbed Einstein equations in the presence of an anisotropic cosmic fluid is an interesting question which is beyond the scope of this work.

 A very interesting question to ask is whether the ideas presented in this work can be used observationally to find the fingerprints of NG and anisotropies in LSS data.  The future LSS surveys such as the  Large Synoptic Survey Telescope (LSST) is designed to obtain photometric redshift for almost 4 billion galaxies. The galaxies are distributed in redshift space  with the distribution peaking around z = 1. This survey  enables us  to determine the galaxy bias
with high accuracy. The galaxy cluster count with combination of other cosmological observations, such as the weak gravitational lensing and CMB data, can measure the full sky bias parameter in the redshift range between 0 to 1, with a
precision as good as $2\%$ accuracy \cite{Zhan:2006gi}. However, the challenge from observational side is that if we want to determine the amplitude of the direction-dependent
bias we need almost a full sky survey with enough statistics in each patch. This is the most important obstacle in determining the bias via galaxy cluster counting. There are other promising observations, such as the 21cm hydrogen intensity power-spectrum, which can be used to detect the fingerprints of NG and anisotropy. It is easier to have a full map data on the hydrogen intensity map although the astrophysical uncertainties will interfere in determining the bias parameter \cite{D'Aloisio:2013sda,Lidz:2013tra,Camera:2013kpa,Mao:2013yaa}. The statistical analysis and future observational forecasts are potential extensions of this work to quantify the chance of the detection. The other point to mention is that the strength of the anisotropy signal depends on $f_{NL}$, the anisotropy parameter $I$ and the angle
$\theta_k$. Now  we have a strong constrain on the local NG from PLANCK data  \cite{Ade:2013ydc}  $f_{NL}=2.7\pm 5.8$ making the signal small.  However there are two crucial points here. First,  there is a  degeneracy between the parameters $f_{NL}$, $I$ and $\theta_k$ and second,  the local NG measured by PLANCK collaboration is on large scales (CMB scale) and there is a room for running of $f_{NL}$ towards the smaller scales (sub CMB/LSS scales) making $f_{NL}$ large enough for our analysis.



\acknowledgments
We would like to thank Razieh Emami for many insightful discussions on anisotropic inflation. We also thank the anonymous referee for the careful and insightful comments on the draft which were very helpful to improve the presentations.


\appendix

\section{Halos bias in the Peak Background Splitting in the context of Excursion Set Theory}
\label{app-EST}

In this appendix first we review the Excursion Set theory (EST) then we apply the Peak Background Splitting (PBS)
in EST context and finally we argue that how we can find the halo bias.
In the context of EST  halo formation can be described as the random walk of
matter density contrast as the smoothing radius goes from very large radius, corresponding
to infinitesimal variance $\sigma ^2$ and small $\delta$, to the scales crossing the linear threshold for collapse $\delta_{c}$ at some finite smoothing radius. This radius is related to the scale in which the halos form.
Within the EST formulation the number density of collapsed objects (dark matter halos of mass $m$ ) per unit mass is given by
\begin{equation}\label{ndensity}
\left(\frac{dn}{dm}\right)=\frac{\bar{\rho}}{m}\partial_{m}\int_{-\infty}^{\delta
_{c}}\Pi_{0}(\delta_{s},\sigma^2_{m},\delta_{c})d\delta_{s},
\end{equation}
where $\sigma^2_{m}$ is the variance of the small scale density
field smoothed with filter (window function) at spatial scale $R$ and $\bar \rho$ is
the background energy density, relating the halo mass to smoothing radius by $m=4\pi\bar{\rho}R^3/3$. Furthermore, $\Pi_{0}(\delta_{s},\sigma^2_{m},\delta_{c})$ is the unconditional probability distribution function (PDF of density fluctuations) of small scale perturbations reaching $\delta_{s}$ (short wavelength density contrast) at
variance $\sigma ^2$. By the subscript $0$ as well as the  unconditional assumption for  PDF of density fluctuations, we mean that the initial condition (first step in random walk) is $\delta_{s}=0$ where $\sigma^2_{m}=0$, and it satisfies the absorbing barrier condition $\Pi_{0}(\delta_{c},\sigma^2_{m},\delta_{c})=0$.

The smoothing procedure will be done by top-hat window function in Fourier space \cite{Weinberg2008}
\begin{equation}
W(x)=\frac{3(\sin x-x \cos x)}{x^3} \, ,
\end{equation}
where $x=kR$, $k$ is the wavenumber and $R$ is Lagrangian radius of collapsed objects, related to the mass via
\begin{equation}
R=\left[\frac{m}{1.162\times10^{12}h^2M_{\odot}\Omega^0_m}\right]^{1/3}Mpc
\end{equation}
in which $m$ is the mass of the structure and $\Omega_m=\Omega^0_m(1+z)^3$.

Since the  variance $\sigma_{m}$ is a monotonic function of mass scale due to matter power spectrum we use $\sigma(M)$ as the 1-D variable in random walk.
So it will be relevant to define a quantity that shows
the probability of first up-crossing in the time $\sigma^2_{m}$ and
$\sigma^2_{m}+d\sigma^2_{m}$ as
\begin{equation}
{\cal{F}}_{0}(\delta_{c},\sigma^2_{m})\equiv -\frac{\partial}{\partial\sigma_m^2}\int_{-\infty}^{\delta_{c}}\Pi_{0}(\delta_{s},\sigma^2_{m},\delta_{c})d\delta_{s} \, .
\end{equation}
Now the number density of collapsed objects will be
\begin{equation}
\left(\frac{dn}{dm}\right)=\frac{\bar{\rho}}{m}|\frac{d\sigma_m^2}{dm}|\times{\cal{F}}_{0}(\delta_c,\sigma_m^2) \, .
\end{equation}
It is worth to mention that the number density of dark matter halos obeys the normalization condition
\begin{equation}
\int(\frac{dn}{dm})mdm=\bar{\rho}
\end{equation}
The key parameter here is the matter density variance which is related to the linear matter power spectrum as
\begin{equation}
\sigma^2(M,z)=\int\frac{d^3k}{(2\pi)^3}P_{L}(k)W^2(kR),
\end{equation}
where the  density contrast power spectrum is \cite{Weinberg2008}
\begin{equation}\label{Eq-powerlinear}
P_{L}(k)=Ak^{n_{s}}T^2(k)D^2(z) \, .
\end{equation}
Here $n_s$ is the spectral index, $A$ is the linear matter primordial power spectrum amplitude in $k=0.002 h^{-1} Mpc$, $T(k)$ is the transfer function and $D(z)$ is the growth factor normalized to scale factor at early times.

The evolution of density contrast is imprinted in  growth function $D(z)$ and  the transfer function $T(k)$,  showing the scale-dependence of gravitational potential during cosmic evolution.  In this work we use the growth function of standard $\Lambda CDM$ cosmology and the transfer function of Bardeen, Bond, Kaiser and Szalay (BBKS) \cite{Bardeen:1985tr} respectively as below
\begin{equation}
D(z)=\frac{5}{2}\Omega_{m}\left[\Omega^{4/7}_{m}-\Omega_{\Lambda}+(1+\frac{\Omega_m}{2})(1+\frac{\Omega_{\Lambda}}{70})\right]^{-1},
\end{equation}
and
\begin{equation}
T(k=q\Omega^0_m h^2 Mpc^{-1})\approx \frac{\ln[1+2.34q]}{2.34q}\times\left[1+3.89q+(16.2q)^2+(5.47q)^3+(6.71q)^4\right]^{-1/4} \, .
\end{equation}
where $\Omega_m=\Omega^0_m a^{-3}/(\Omega^0_ma^{-3}+\Omega_{\Lambda})$ and  $\Omega_{\Lambda}=\Omega^0_{\Lambda}/(\Omega^0_m a^{-3}+\Omega_{\Lambda} )$.
We define the growth function, in a way that is normalized to scale factor in deep matter dominated era.
An important point to note is that the variance is linearly dependent to  growth function due to Eq. (\ref{Eq-powerlinear}) where $\sigma^2(M,z)=\sigma^2(M,z=0)[D(z)/D(z=0)]^2$, where
$\sigma^2(M,z=0)$ is the present value of variance. This redshift-dependance is important in the sense that the statistics of structure and even the bias parameter will be redshift-dependent. In Fig. (\ref{biasz}) we showed the redshift-dependence of bias parameter for our specific anisotropic inflation model. In the matter dominated era $D(z)$ scales like $1/(1+z)$ which is a decreasing
function with respect to redshift.
Now that we set the matter density variance  and large scale statistics of matter distribution, by setting the initial
condition of perturbations we can find the mass function of structures in the Universe.
For the Gaussian initial condition the probability distribution function (PDF of density fluctuations) in Fourier-space top-hat filter is \cite{Bond1991}
\begin{equation}
\Pi_{0}(\delta_{s},\sigma^2_{m},\delta_{c})={\cal{P}}_{G}(\delta_{s},\sigma^2_{m})-{\cal{P}}_{G}(2\delta_{c}-\delta_{s},\sigma^2_{m}),
\end{equation}
where ${\cal{P}}_{G}$ is the Gaussian  PDF of density fluctuations.

 The assumption of the universality of mass-function (like Press-Schechter) yields
 \begin{equation}
 \Pi_{0}(\delta_{s},\sigma_{m}^2,\delta_{c})=F(\frac{\delta_{s}}{\sigma_{m}},\frac{\delta_{c}}{\sigma_{m}}),
 \end{equation}
which leads us to write $\sigma^2_{m}{\cal{F}}_{0}$ just as a function of
 threshold quantity $\nu=\delta_{c}/\sigma_{m}$
 \begin{equation}
 \sigma^2_{m}{\cal{F}}_{0}(\delta_{c},\sigma^2_{m})=\frac{\nu f(\nu)}{2},
 \end{equation}
 where $f(\nu)$ is the usual Gaussian factor in Press-Schechter theory \cite{Press:1973iz}. Now in order to find the halo bias in Gaussian and non-Gaussian inflationary models, first
 we discuss the PBS. We will show how this method is applicable for non-Gaussian fields.

\label{app-PBS}
For inflationary models with primordial non-Gaussianity, the gravitational potential is usually written as  \cite{Salopek1990,Salopek:1990jq,Komatsu:2001rj}
\begin{equation}
\Phi(x)=\phi(x)+f_{NL}([\phi(x)]^2-\langle\phi^2\rangle)
\end{equation}
where $\phi$ is a Gaussian random field. The above relation is the simplest extension of Bardeen potential to include non-linearity, called local non-Gaussianity. We can generalize the local type non-Gaussianity to arbitrary shape, simply by replacing the non-linear term by a kernel \cite{Scoccimarro:2011pz}
\begin{equation}
\Phi(x)=\phi(x)+f_{NL}K[\phi(x),\phi(x)] \, .
\end{equation}
In order to obtain the bias parameter, we can use the PBS idea on the Gaussian potential $\phi=\phi_{s}+\phi_{l}$, where $\phi_{l}$ is the long-wavelength mode of the potential and $\phi_s$ is the short-wavelength mode. Since the large scale perturbations change the background for small scale modes, the  PDF of density fluctuations for small scale perturbations is modified to the conditional  PDF of density fluctuations, $\Pi(\delta_s, \sigma_m^2, \delta_c;\delta_l,\sigma_l^2)$. In conditional PDF of density fluctuations, the initial condition is $\sigma_m \to \sigma_l$ in large scale limit (first step in random walk), in contrast with the unconditional case in which the variance vanishes on large scales.

One of the important assumptions in PBS  method is the Markovianity where we can write
\begin{equation}
\Pi(\delta_s,\sigma^2_m,\delta_c;\delta_l,0)\approx \Pi(\delta_s-\delta_l,\sigma^2_m,\delta_c-\delta_l)
\end{equation}
where $Pi$ is the PDF of density perturbations. The Markovianity condition let us to change the threshold of critical density and density fluctuation of the structure by $\delta_l$. In this work we are in he regime  where we can neglect the non-Markovianity induced from  the primordial NG (for an extended discussion refer to Scoccimarro et al. \cite{Scoccimarro:2011pz}).

On the other hand, the existence of any type of NG modifies the  PDF of density fluctuations in a way that  it is no longer independent of higher cumulants, while in Gaussian case the  PDF of density fluctuations is completely determined by zeroth order ($\delta_s$) and first order ($\sigma_m^2$) cumulants. By the splitting procedure explained above, and knowing that the  PDF of density fluctuations is now a function of all cumulants, we can expand the Lagrangian halo over-density as an expansion over large-scale $\phi_l$ modes \cite{Scoccimarro:2011pz},
\begin{equation}\label{Eq-halobiasEPS}
\delta^{L}_{h}=\int d^3k\frac{\partial _{m}\int _{-\infty}^{\delta_{c}}d\delta_{s}({D\Pi}/{D\phi _{l}})_{0}}{\partial _{m}\int _{-\infty}^{\delta_{c}}d\delta_{s}\Pi_{0}(\delta_{s},\sigma^2,\delta_{c})}\phi_{l}+\frac{1}{2}
\int\int d^3k_1d^3k_2\frac{\partial _{m}\int _{-\infty}^{\delta_{c}}d\delta_{s}({D^2\Pi}/{D\phi _{l} D\phi_{l}})_{0}}{\partial _{m}\int _{-\infty}^{\delta_{c}}d\delta_{s}\Pi_{0}(\delta_{s},\sigma^2,\delta_{c})}\times \phi_{l}(\vec{k}_{1})\phi_{l}(\vec{k}_{2})+...
\end{equation}
where $(...)_{0}$ means that the corresponding quantity is evaluated at $\phi_{l}=0$. Note that because the background is now modified by large scale perturbation $\phi_l$, the cumulants are now functions of $\phi_l$. The first derivative in Eq. (\ref{Eq-halobiasEPS}), to all orders in primordial NG, is
\begin{equation}\label{Pdif}
(\frac{D\Pi}{D\phi_{l}(\vec{k})})_0=\sum_{p=1}^{\infty}\left(\frac{\partial\Pi}{\partial c^{(p)}}\right)_0\left(\frac{D c^{(p)}}{D \phi_{l}(k)}\right)_0=\left(\frac{\partial \Pi}{\partial \delta _{l}}\right)_0\left(\frac{D \delta_{l}}{D\phi_{l}(\vec{k})}\right)+\sum_{p=2}^{\infty}\frac{\partial \Pi_{0}}{\partial c_{m}^{(p)}}\left(\frac{D c^{(p)}}{D\phi_{l}(\vec{k})}\right)_{0} \, ,
\end{equation}
where the cumulants are defined by
\begin{equation}
c^{(1)}\equiv \delta_{l}, ~~~  c^{(2)}_m\equiv\sigma_m^2, ~~~ c^{(2)}\equiv\sigma^2(\phi_{l}), ~~~ c^{(p)}_m\equiv {\langle \delta ^p_s\rangle}_c, ~~~ c^{(p)}\equiv {\langle \delta ^p_s(\phi_{l})\rangle}_c
\end{equation}
in which the subscript $m$ for cumulants show that they are evaluated in the absence of the background $\phi_l$ (independent of  $\phi_l$).  It is worth to indicate that to first order in $f_{NL}$, only the first two terms in the Taylor expansion above ($p=1,2$) contributes to the bias, while $p=3$ contributes to ${\cal{O}}(f^2_{NL})$ and ${\cal{O}}(g_{NL})$.

Now we can obtain the bias parameter, using the above formulation.
The $p=1$ contribution is the usual scale-independent bias presented in initially Gaussian case.
Keeping in mind $b\equiv\delta_{h}/\delta_l$, we have
\begin{equation}
\label{b1l}
p=1: ~~ b_{1L}=\frac{\partial_m \int(\partial\Pi/\partial\delta_l)_0}{\partial _m\int\Pi_0}=\left[\frac{\partial}{\partial\delta_{l}}\ln(\frac{dn(\delta_{l})}{d\ln{m}})\right],
\end{equation}
which can be simplifies as
\begin{equation}
b_{1L}=\frac{\partial}{\partial\delta_l}\ln(n(\delta_l)) \, .
\end{equation}

The $p=2$ contribution is the scale-dependent correction to the leading order bias coming from
primordial NG
\begin{equation}
p=2: ~~ b_{2L}=\frac{\partial_m [I_{21}\int \partial\Pi_0/\partial\sigma_m^2]}{M(k)\partial _m\int\Pi_0} \, .
\end{equation}
The quantity $I_{21}$ includes the information about the primordial NG which is the derivative of second cumulant $\sigma_m^2$, (p=2), with respect to
long wavelength mode $\phi_l$  which is obtained to be \cite{Scoccimarro:2011pz}:
\begin{equation}
\label{I21-app}
I_{21}(k,m)=\frac{1}{P_{\phi}(k)}\int B_{\hat{\delta}\hat{\delta}\phi}(q,k-q,-k)d^3q,
\end{equation}
where $B_{\hat{\delta}\hat{\delta}\phi}$ is the cross bispectrum of small-scale smoothed density $\hat{\delta}$ and $\phi$.

So far only the Lagrangian bias appeared in our analysis because the
peaks are those of the initial density field (linearly extrapolated). By the
standard assumptions that halos move coherently with the underlying dark matter,
and using the techniques outlined in \cite{Mo:1995cs,Catelan:1997qw,Efstathiou:1988tk,Cole:1989vx}, one can obtain the final Eulerian bias as
\begin{equation}\label{app-eq-Eubia}
b_{E}=1+b_{1L}+b_{2L} \, ,
\end{equation}
Note that, due to the existence of primordial NG, the leading order
scale-independent bias also modifies as
\begin{equation}\label{app-b1l}
b_{1L}=\frac{2}{\delta_c}\partial_{\ln\sigma_m^2}\ln(\sigma_m^2{\cal{F}})=b_{1L(G)}+b_{1L(NG)},
\end{equation}
where in Eq. \eqref{app-b1l} we have omitted the subscript of ${\cal{F}}$, which means that the NG will change the mass function, resulting in a modification of $b_{1L}$. As a result we have
\begin{equation}
b_{1L(G)}=\frac{2}{\delta_c}\partial_{\ln\sigma_m^2}\ln(\sigma_m^2{\cal{F}}_{0})
\end{equation}
and
\begin{equation}
b_{1L(NG)}=\frac{2}{\delta_c}\partial_{\ln\sigma_m^2}\ln(\sigma_m^2{\cal{R}}_{NG})=\frac{\partial \ln {\cal{R}}_{NG}(m,f_{NL})}{\partial\delta_l},
\end{equation}
{{as described in Section (\ref{SecBias}). }}
%
\begin{equation}
\label{RNG2}
{\cal{R}}_{NG}(m,f_{NL})=1+\frac{1}{6}x(x^2-3)s_{3}(x)-\frac{1}{6}(x-1/x)\frac{ds_3(x)}{d\ln(x)},
\end{equation}
where $x\equiv{\delta_c}/{\sigma_M}$, $\delta_c=1.68$ is the critical density and $s_{3}$ is the reduced skewness defined as
\begin{equation}
s_{3}(R)\equiv\frac{\langle\delta^3_R\rangle}{\langle\delta^2_R\rangle ^{3/2}}=\frac{\langle\delta^3_R\rangle}{\sigma^3_m} \, .
\end{equation}
The skewness is related to the matter bispectrum as
\begin{equation}
\langle\delta_R^3\rangle=\int \frac{d^3q_1}{(2\pi)^3}\frac{d^3q_2}{(2\pi)^3}W(Rq_1)W(Rq_2)W(Rq_{12})M(q_1,z)M(q_2,z)M(q_{12},z)B_0(q_1,q_2,q_{12}),
\end{equation}
where  $\vec{q}_{12}=-(\vec{q}_{1}+\vec{q}_{2})$ and $W(kR)$ is the window function in Fourier space, smoothing perturbations up to scale $R$.
In obtaining the  mass function of non-Gaussianity model in this approximation, we have assumed that all the deviation is imprinted in the skewness which may not be entirely true. In order to
improve  the results, numerical simulations are done \cite{Grossi:2009an,Maggiore:2009rx,Paranjape:2011uk}. Consequently, a scaling parameter $\kappa$ defined by $R_{NG}(x)\rightarrow R_{NG}(\kappa x)$ are introduced where, in the work of
\cite{Sefusatti:2011gt}, from simulation of \cite{Desjacques:2008vf}, it  is obtained to be $ \kappa = 0.91$. (For a similar correction from simulation see \cite{Tinker:2008ff}).


On the other hand, the scale-dependent bias can be rewritten as
\begin{equation}
b_{2L}=\frac{I_{21}(k,m)}{2\sigma_m^2 M(k,z)}\delta_c b_{1L}+\frac{1}{M(k,z)}\partial_{\ln\sigma_m^2}(\frac{I_{21}(k,m)}{\sigma_m^2}) \, .
\end{equation}
So the total Eulerian bias up to first order in $f_{NL}$ can be split to scale-independent $b_{si}$ and scale-dependent $b_{sd}$ terms as
\begin{equation}
b_{t}=b_{si}+b_{sd}
\end{equation}
where $b_{si}$ and $b_{sd}$ are
\begin{equation}
b_{si}\equiv b_{G}+b_{1L(NG)},
\end{equation}
with $b_{G}\equiv 1+b_{1L(G)}$ and
\begin{equation}
b_{sd}=b_{2L}+{\cal{O}}(f^2_{NL})+{\cal{O}}(g_{NL})+... \, .
\end{equation}

\end{document}